\documentclass[a4paper,fleqn]{cas-dc}

\usepackage[authoryear]{natbib}
\usepackage[authoryear]{natbib} % Use natbib with author-year format
\usepackage{xcolor}             % For text coloring



\usepackage{graphicx,times}
\usepackage{subcaption}
\usepackage{amssymb,amsmath}
\usepackage{aas_macros}
\usepackage{hyperref}
\usepackage{soul}
\let\oldAA\AA
\renewcommand{\AA}{\text{\normalfont\oldAA}}
\usepackage{pdflscape}

\def\tsc#1{\csdef{#1}{\textsc{\lowercase{#1}}\xspace}}
\tsc{WGM}
\tsc{QE}
\begin{document}

%%%%%%%%%%%%%%%%%%%%%%%%%%%%%%%%%%%%%%%%%%%%%%%
\let\WriteBookmarks\relax
\def\floatpagepagefraction{1}
\def\textpagefraction{.001}

% Short title
\shorttitle{Multi-wavelength analysis of FSRQ B2\,1348+30B: Constraints on the jet power}    

% Short author
\shortauthors{}

% Main title of the paper
%\title [mode = title]{Multi-wavelength temporal and spectral study of the FSRQ B2\,1348+30B}  
\title [mode = title]{Multi-wavelength analysis of FSRQ B2\,1348+30B: Constraints on the jet power}  

% Title footnote mark
% eg: \tnotemark[1]
\tnotemark[<tnote number>] 

% Title footnote 1.
% eg: \tnotetext[1]{Title footnote text}
%\tnotetext[<tnote number>]{<tnote text>} 

% First author
%
% Options: Use if required
% eg: \author[1,3]{Author Name}[type=editor,
%       style=chinese,
%       auid=000,
%       bioid=1,
%       prefix=Sir,
%       orcid=0000-0000-0000-0000,
%       facebook=<facebook id>,
%       twitter=<twitter id>,
%       linkedin=<linkedin id>,
%       gplus=<gplus id>]

\author[1]{Sajad Ahanger}

% Corresponding author indication
\cormark[1]

% Footnote of the first author
%\fnmark[<footnote mark no>]

% Email id of the first author
\ead{sajadphysics21@gmail.com}

% URL of the first author
%\ead[url]{<URL>}

% Credit authorship
% eg: \credit{Conceptualization of this study, Methodology, Software}
%\credit{<Credit authorship details>}

% Address/affiliation
\affiliation[1]{organization={Department of Physics, University of Kashmir},
%            addressline={}, 
            city={Srinagar},
%          citysep={}, % Uncomment if no comma needed between city and postcode
            postcode={19006}, 
       %    state={},
            country={India}}

\author[2,3]{Sunder Sahayanathan}

% Footnote of the second author
%\fnmark[]

% Email id of the second author2009ApJ...692...32D
\ead{sunder@barc.gov.in}

\affiliation[2]{organization={Astrophysical Sciences Division, Bhabha Atomic Research Centre},
%            addressline={}, 
            city={Mumbai},
%          citysep={}, % Uncomment if no comma needed between city and postcode
            postcode={400085}, 
          %  state={Maharashtra},
            country={India}}
            
 \affiliation[3]{organization={Homi Bhabha National Institute},
%            addressline={}, 
            city={Mumbai},
%          citysep={}, % Uncomment if no comma needed between city and postcode
            postcode={400094}, 
            %state={Maharashtra},
            country={India}}   
            
%\cormark[1]     
% URL of the second author
%\ead[url]{}

% Corresponding author indication

% Credit authorship
%\credit{}

\author[5]{Sitha K. Jagan}
%\ead{sithajagan@gmail.com}

\affiliation[5]{organization={Artificial Intelligence Research and Intelligent Systems},
            city={Thelliyoor},
            postcode={689644}, 
            country={India}}

\credit{}
\author[4]{Shah Zahir}
\ead{shahzahir4@gmail.com}
\affiliation[4]{organization={Department of Physics, Central University of Kashmir},
%            addressline={}, 
            city={Ganderbal},
%          citysep={}, % Uncomment if no comma needed between city and postcode
            postcode={191201}, 
       %    state={},
            country={India}}

\author[1]{Naseer Iqbal}

% Corresponding author indication
%\cormark[5]

% Footnote of the first author
%\fnmark[<footnote mark no>]

% Email id of the first author
%\ead{sajadphysics21@gmail.com}

% URL of the first author
%\ead[url]{<URL>}
\credit{}

\cortext[1]{Corresponding author}
% Credit authorship
% eg: \credit{Conceptualization of this study, Methodology, Software}
%\credit{<Credit authorship details>}

% Address/affiliation

% Footnote text
%\fntext[1]{}

% Abstract of the paper
\begin{abstract}
%The striking variability and flaring activity observed in the flat-spectrum radio quasar (FSRQ) B2\,1348+30B provide valuable insights into the dynamics and energetics of relativistic jets.  
We present 14.5-year multi-wavelength analysis of flat-spectrum radio quasar  B2\,1348+30B using Swift-UVOT, Swift-XRT, and \emph{Fermi}-LAT 
observations. In the $\gamma$-ray band, the 3\,day bin lightcurve reveals two major flaring events on 2010-09-19 (55458 MJD) and 2022-05-26 (59725 MJD)  detected 
at flux levels $(2.5\pm 0.5) \times 10^{-7}\,\rm{ph\,cm^{-2}\,s^{-1}}$ and $(5.2\pm 0.6) \times 10^{-7}\,\rm{ph\,cm^{-2}\,s^{-1}}$. The Bayesian block analysis of the flares suggested the variability timescale to be $\leq$ 3\,day. To study the dynamic nature of the source, multi-wavelength spectrum was obtained for three flux states which includes the two
flaring state and a relative low state. The $\gamma$-ray spectra of the source in all the states are well fitted by a power-law model
with maximum photon energy < 20 GeV. In X-ray, a power-law model can explain the flaring state spectra while a broken power-law with 
extremely hard high energy component was required to model the low flux state. This indicates the presence of the low energy cutoff in the 
Compton spectral component. A simple one-zone leptonic model involving synchrotron, synchrotron self Compton and external Compton 
mechanism can successfully reproduce the broad-band spectral energy distribution of all the flux states. The model parameters suggest
significant increase in the jet Lorentz factor during the high flux states. Further, the best fit parameters are used to constrain 
the minimum energy of the emitting electron distribution from the hard high energy spectrum of the low flux state. This analysis 
was extended to draw limits on the kinetic power of the blazar jet and was compared with the Eddington luminosity of the central
black hole.

%of $(0.9\pm0.1) \times 10^{-6} ph\,cm^{-2}\,s^{-1}$ observed on 2022 May 26. 
%Using the Bayesian block algorithm, the estimated variability timescale $\leq$ 3\,day constrains the $\gamma$-ray emission region size to $4.3 \times 10^{16}\,cm$. The $\gamma$-ray spectra of the source in all the states are well fitted by a power-law model, and no photons with energy exceeding 20 GeV and significance  $\geq3\sigma$ were detected. The UVOT spectral analysis, along with variability studies, suggests a possible disk component contributing to the low-energy emission from the source. A one-zone leptonic model, incorporating synchrotron, synchrotron-self-Compton (SSC), and external-Compton (EC) processes, successfully reproduces the broad-band SED for each flux state. 
%The analysis shows that the bulk Lorentz factor nearly doubles during flaring states, suggesting strong shocks from high-speed ejecta near the black hole drive the flares. Utilising the extremely hard X-ray spectrum in state S2, we constrain the minimum particle energy cut-off to approximately 4 and the jet power to lie within 46.41 for a light jet and 48.62 for a heavy jet. This comprehensive study highlights the connection between flaring activity and jet energetics, offering critical insights into particle acceleration and the physical conditions governing blazar emissions.
\end{abstract}


\maketitle

%%%%%%%%%%%%%%%%% BODY OF PAPER %%%%%%%%%%%%%%%%%%
\section{INTRODUCTION}
\label{intro}
Active galactic nuclei (AGN) are one of the most energetic and luminous objects in the extragalactic universe, powered by supermassive 
black holes with masses ranging from  $10^{6}-10^{10}$$M_\odot$. Blazars are a subclass of these AGN characterised by  relativistic jet of 
matter directed towards our line of sight where the constituent particles undergo continuous acceleration \citep{1984RvMP...56..255B,1995PASP..107..803U}. 
Hence, the emission from  the blazar jet is highly Doppler boosted that it outshines the host galaxy. Its spectrum is predominantly non-thermal in nature extending 
across radio, optical/UV to high energy [HE >100\,MeV] or 
very high energy [VHE >100\,GeV] $\gamma$-ray energies \citep{2009ApJ...692...32D}.
Besides these, blazars also exhibit rapid flux variability on timescales ranging from minutes to days 
and variable polarization with significant changes in polarization angle 
%({\bf refer Athira paper also})
\citep{2015MNRAS.453.1669B,2016A&A...590A..10K,2018Galax...6...34G,2024JApA...45...35B}. 
%({\bf This line is repetition: Removed now})
Based on their optical spectra, blazars are classified into flat-spectrum radio quasars (FSRQs), which exhibit
broad emission/absorption line features (EW > 5\AA), and BL Lacertae objects (BL Lacs), which lack emission/absorption 
lines (EW < 5\AA) in their optical spectra and are often featureless \citep{1991ApJ...374..431S,1991ApJS...76..813S,1995PASP..107..803U,1997A&A...325..109S}.

%%%%%%%%%%%%%%%%%%%%%%%%%%%%%%%%%%%%%%%%%%%%%%%%%%%%%%%%%%%%%%%%%%%
The broad-band spectral energy distribution (SED) of blazars is characterised by two prominent peaks, with the low energy component extending from radio 
to soft X-ray and is well understood to be synchrotron emission from a relativistic non-thermal electron distribution losing its energy in the 
jet magnetic field \citep{1978bllo.conf..328B,ghisellini1989}. The
high energy component, spanning from hard X-rays to MeV/GeV $\gamma$-rays, is often debated under leptonic or hadronic (or both) emission scenarios. In the leptonic 
scenario, the high energy emission is  interpreted as the inverse Compton (IC) scattering of low-energy target photons by the relativistic electron
distribution in the jet. The target photons can be synchrotron emission itself, in this case the IC scattering process is referred as synchrotron 
self-Compton (SSC) \citep{1985A&A...146..204G,1996ApJ...461..657B} or the photon field external to the jet (accretion disk, broad line region (BLR) or 
dusty torus) and the IC scattering process is called as the external Compton 
(EC) \citep{1993ApJ...416..458D,1994ApJ...421..153S,2018MNRAS.477.4749C}.
The hadronic models, on the other hand, interpret the high energy component of SED through 
proton-synchrotron \citep{2000NewA....5..377A,2001APh....15..121M,2016ApJ...826...54D,2016ApJ...817...61P,Das:2021xcd} or nuclear cascades initiated by 
proton-photon \citep{1992A&A...253L..21M,1999MNRAS.302..373B,2002PhRvD..66l3003N,2023arXiv231003561S} or proton-proton 
interactions \citep{1993PhRvD..47.5270N,2009PhRvD..80h3008N,2013ApJ...768...54B,2017A&A...603A.135N,2023ApJ...948...75M}. 
Hadronic interactions will be associated with the production of neutrinos and hence the detection of neutrinos from blazars
favours these models \citep{2017AIPC.1792e0027C,2024ApJ...977...42R}. However, this interpretation has difficulty in explaining the observed rapid variability in blazars  \citep{1998Sci...279..676B,2013EPJWC..6105009W,inproceedings2017}. %{\bf (Add Buckley Science paper)}.

Despite significant progress in our understanding on the emission processes, the kinetic power of the blazar jet (or the AGN jet in general)
is largely an open question. Modelling the non-thermal emission and variability studies are quite successful in providing estimates for flow velocity 
\citep{1995MNRAS.273..583D,1998ApJ...509..608T,2022A&A...658A.173T}; however, jet power estimation demands the nature and number of its matter species (lepton and/or hadron). 
Under leptonic emission scenario, limits of blazar jet power can be 
estimated by assuming the jet matter to be only made of electrons and positrons (light jet) or electron-proton plasma with proton number equal to that of
non-thermal electron (heavy jet). In the latter case, the protons are also assumed to be cold and do not take part in the radiative processes 
%({\bf more refernces!Celloti ghisellini paper, sitha's paper.}) 
\citep{2008MNRAS.385..283C,2013ApJ...768...54B,2021MNRAS.506.3996J}.
Being less massive, the light jet case will lack the necessary inertia to launch the jet up to kilo/mega parsec scales. The heavy jet, on the other hand,  
estimates jet powers which are greater than (or at least equal to) the accretion power \citep{2014Natur.515..376G}. Hence, the actual jet power
is expected to lie within these limits and probably involve electron-positron-proton plasma. The jet power estimated from the hadronic emission scenario
is also extremely large compared to the accretion power and poses severe problems in accounting for the energy budget \citep{2015MNRAS.450L..21Z}.
A vital quantity to estimate the jet power (at least under leptonic emission scenario) is the lepton number density which is largely decided by the 
choice of minimum available non-thermal electron energy. However, estimation of this from the observations is challenging and one can obtain limiting 
values based on the spectral curvature \citep{2016ApJ...831..142M,2021MNRAS.506.3996J}.

Blazar B2\,1348+30B (or 4FGL J1350.8+3033) is a FSRQ located at a redshift of $z \sim 0.712$ \citep{2015MNRAS.452.4153A} 
with coordinates R.A = 207.714798 and Dec = 30.558800 \citep{2011AJ....142...89P,2023MNRAS.519.6349D}. 
%observed  $\gamma$-ray luminosity of $\sim 3.01\times10^{46}\,erg\,s^{-1}$, 
The source was reported in the SDSS DR7 quasar catalog with broad emission lines of $\rm H\beta$ and $\rm Mg\,II$ \citep{2015yCat..22190001O} and the 
estimated black hole mass  $\sim 1.86\times10^{08}M_\odot$ \citep{2021ApJ...913...93C}. 
%and shows a strong link between the jet kinetic power and spin of the black hole \citep{2014Natur.515..376G,2012ApJ...757...25L,2021ApJ...913...93C}. 
The all-sky variability analysis conducted by \emph{Fermi} has recorded a high-energy flare of the source in 2013 
with a maximum $\gamma$-ray luminosity $\sim 3.01\times10^{46}\,erg\,s^{-1}$ \citep{2013ApJ...771...57A,2021ApJ...913...93C}. 
%Furthermore, the source has been also reported in 2022 by the two  astronomical telegrams. 
The source was again observed to be in an elevated $\gamma$-ray state on 2022 May 26 and 27, with a 
peak daily averaged $\gamma$-ray flux $(E >100MeV)$ of $(0.9\pm0.1) \times 10^{-6} ph\,cm^{-2} s^{-1}$ \citep{2022ATel15402....1V}.
This flux level equals to $\sim 60$ times the average flux recorded in the fourth \emph{Fermi}-LAT (4FGL) catalog  \citep{2020ApJS..247...33A} and
was the highest LAT daily flux ever reported for this source. Contemporaneous with this $\gamma$-ray flare, the source was also observed by 
IceCube in search of neutrino events, albeit no significant detection and only upper limits were reported \citep{2022ATel15439....1T}. 
%It was also reported by \citep{2022ATel15439....1T} on 2022 June 16, that IceCube had performed 
%a search for track-like muon neutrino events arriving from the direction of the source during its $\gamma$-ray flare using a time window of 61 days which 
%only resulted in an upper limit detection. 
In this paper, we perform a long-term muti-wavelength analysis of the FSRQ B2\,1348+30B using \emph{Fermi}-LAT, Swift-XRT and Swift-UVOT observations.
We reproduce the broad-band SED of the source using synchrotron and inverse Compton emission processes and also investigate the minimum available electron
energy and the jet power. The paper is organised as follows: 
section \S\,\ref{sec:data_anly} elaborates on the specifics of the multi-wavelength data and the methodologies used for analysis. The results of the temporal 
and spectral analysis carried out at various wavelengths are discussed in section \S\,\ref{sec:results}. Finally, sections \S\,\ref{sec:sum}
concludes with the summary of the work. Throughout this work, we have adopted a cosmology with $\Omega_M = 0.3$, $\Omega_\Lambda = 0.7$, 
and $H_0 = 71 \, \text{km s}^{-1} \, \text{Mpc}^{-1}$.

%%%%%%%%%%%%%%%%%%%%%%%%%%%%%%%%%%%%%%%%%%%%%%%%%%%%%%%%%%%%%%%
\begin{table*}
\centering
\renewcommand{\arraystretch}{1.2} 
\begin{tabular}{l  c c c c}
\hline 
\hline
Observation ID & Time (MJD) & Time (Gregorian) & XRT exposure (ks) \\
\hline

00039177001 & 55442.9 & 2010-11-10 & 1.25 \\
00039177002 & 55444.3 & 2010-11-11 & 1.59 \\
00039177003 & 55528.8 & 2011-02-03 & 2.11 \\
00013619001 & 59049.6 & 2020-10-06 & 2.98 \\
00013619002 & 59052.3 & 2020-10-09 & 2.90 \\
00013619003 & 59055.1 & 2020-10-12 & 3.13 \\
00013619004 & 59058.7 & 2020-10-15 & 2.13 \\
00013619005 & 59061.1 & 2020-10-18 & 1.63 \\
00013619006 & 59064.3 & 2020-10-21 & 1.18 \\
00046508002 & 59156.2 & 2021-01-21 & 7.87 \\
00039178002 & 59729.8 & 2022-09-07 & 1.77 \\
00039178003 & 59731.8 & 2022-09-09 & 1.93 \\
00039178004 & 59732.2 & 2022-09-09 & 5.43 \\
00039178005 & 59733.5 & 2022-09-10 & 6.06 \\
00039178006 & 59734.3 & 2022-09-11 & 3.00 \\
00039178007 & 59743.4 & 2022-09-20 & 8.97 \\
00039178009 & 59752.5 & 2022-09-29 & 2.85 \\
00039178010 & 59755.1 & 2022-10-02 & 9.02 \\
00039178011 & 59764.5 & 2022-10-11 & 8.72 \\
\hline
\hline
\end{tabular}
%\vspace{0.5cm}
\caption{Details of Swift-XRT observations used in this work.}
\label{tab:SUV}
\end{table*}
%%%%%%%%%%%%%%%%%%%%%%%%%%%%%%%%%%%%%%%%%%%%%%%%%%%%%%%%%%%%%%%%%%
\section{OBSERVATIONS AND DATA ANALYSIS}
\label{sec:data_anly}

To investigate the temporal and spectral characteristics of the FSRQ B2\,1348+30B during both low and high flux periods, we utilised UV/optical data from Swift-UVOT, X-ray data from Swift-XRT and $\gamma$-ray data from \emph{Fermi}-LAT instruments. The detailed analysis procedures for these datasets are outlined below.\\

\subsection{\emph{Fermi}-LAT ${\gamma}$-ray Analysis}
\emph{Fermi}-LAT (\emph{Fermi}-Large Area Telescope) is a pair-conversion $\gamma$-ray telescope, one of the two scientific instruments on board the \emph{Fermi} $\gamma$-ray Space Telescope \citep{2009ApJ...697.1071A}. \emph{Fermi}-LAT was launched on 2008 June 11, and the energy range covered by it extends from 
%It was launched on 2008 June 11 with an energy range of 0.1 -- 100\,GeV although its sensitivity extends from 
20\,MeV to $>$ 1\,TeV \citep{2017ApJS..232...18A}. The telescope has an orbital period of $\sim$ 96 mins with a field of view (FoV) of 2.4\,sr and scans the entire sky in $\sim$3 hrs. Since its launch in 2008, FSRQ\,B2 1348+30B has been continuously monitored.

In this study, we have analysed more than 14 years of \emph{Fermi}-LAT PASS 8 data from 2008-08-04 to 2022-11-03 (54682–59886 MJD). 
%The analysis was conducted using the latest version of the Python package 
The analysis was conducted utilizing 
\emph{Fermi}py--v1.1.4 \citep{2017ICRC...35..824W} and \emph{Fermitools--v2.2.0}\footnote{\url{https://fermi.gsfc.nasa.gov/ssc/data/analysis/documentation/}} with a  $15^\circ$ region of interest (ROI) centered around the source. We used \emph{‘evclass=128’} and \emph{‘evtype=3’} to analyse the data. Here \emph{‘evclass=128’} determines event quality based on background rejection, and ‘evtype=3’ specifies the reconstruction type for angular and energy resolution. The $\gamma$-ray analysis was performed with a binned likelihood method over the energy range of 0.1–300\,GeV, utilizing the latest instrument response function (IRF) \emph{"P8R3\_SOURCE\_V3"}.
%\textbf{The $\gamma$-ray analysis was performed within the energy range of 0.1--300\,GeV, utilising the latest instrument response function (IRF) \emph{"P8R3\_SOURCE\_V3"} and employing a binned likelihood method.}
To minimize contamination from $\gamma$-rays due to interaction in the Earth's atmosphere, a $90^\circ$ zenith angle cut was applied. Utilising the \emph{Fermi}-LAT 4FGL catalog \citep{2020ApJS..247...33A}, we generated an XML model file encompassing all sources within the ROI. During the  analysis, the source parameters were left free  within the ROI and constrained to their 4FGL catalog values beyond ROI. To account for diffuse background emission, we included the Galactic diffuse emission model \emph{"gll\_iem\_v07.fits"} and an extragalactic isotropic emission component \emph{"iso\_P8R3\_SOURCE\_V3\_v1.txt"} in the model file during the fitting process. The detection significance of each source within the ROI was assessed by utilizing the Test Statistics (TS), defined as TS = 2\,$\Delta$\,log(likelihood), comparing models with source and excluding the source \citep{1996ApJ...461..396M}. In this work, we generated $\gamma$-ray lightcurves with different time binning and SEDs for three selected epochs, respectively.

%%%%%%%%%%%%%%%%%%%%%%%%%%%%%%%%%%%%%%%%%%%%%%%%%%%%%%%%%%%%%%%%%%%%
\subsection{Swift-XRT/UVOT Analysis}
\emph{Neil Gehrels Swift Observatory} \citep{2004ApJ...611.1005G} is a space based telescope equipped with instruments such as the X-Ray Telescope (XRT), 
Ultraviolet/Optical Telescope (UVOT) and X-ray Burst Alert Telescope (BAT). 
%This configuration provides the capability to explore the cosmos in different wavebands simultaneously. 
In this study, we analysed all Swift observations of FSRQ B2\,1348+30B available during the \emph{Fermi} observation of the source and the details are 
provided in Table\,\ref{tab:SUV}.
%%%%%%%%%%%%%%%%%%%%%%%%%%%%%%%

The Swift-XRT \citep{2005SSRv..120..165B} data, taken in photon counting (PC) mode, were processed using the \emph{"xrtpipeline"}, Version: 0.13.7 and calibration ﬁle (CALDB, version: 20190910) in the energy range from 0.3 to 10.0\,keV. The \emph{"xselect"} tool is used to extract a circular source region of 20 pixels and a  circular background region of 50 pixels away from the source location for the extraction of the spectrum. Further, \emph{"xrtmkarf"} and \emph{"grppha"} are used to create the ancillary response ﬁle (ARF) and group the spectra respectively, such that each bin contains at least 20 counts. The grouped spectra are then fitted in the \emph{"XSPEC (version - 12.13.0)"} \citep{1996ASPC..101...17A} which is built into the \emph{HEASOFT} package. We used \emph{tbabs}*power-law (PL), \emph{tbabs}*log-parabola (LP), and \emph{tbabs}*broken power-law (BPL) models to fit the spectra. 
The tbabs component represents Galactic absorption in XSPEC and accounts for the X-ray absorption by the interstellar medium.
 During the spectral fit, the value of pivot energy was fixed at 1\,keV. The fitting parameters are provided in Table\,\ref{tab:x-ray}. To account for galactic absorption, a neutral hydrogen column density ($n_H$) of $1.44\times10^{20}\,cm^{-2}$ \citep{2005yCat.8076....0K} was used, as given in the \emph{HEASARC} webpage. 
%\vspace{0.5mm}\\
%%%%%%%%%%%%%%%%%%%%%%%%%%%%%%%%%%%%%%%%%%%%%%%%%%%%%%%%%%%%%%%%%%%%
\begin{table*}
\centering
\renewcommand{\arraystretch}{1.6} 
  \begin{tabular}{ccccccccccc}

    \hline
    \hline
  & \multicolumn{1}{c}{PL} & \multicolumn{2}{c}{LP} & \multicolumn{3}{c}{BPL}& \multicolumn{1}{c}{PL, LP, BPL} & \multicolumn{2}{c}{LP, BPL} \\
    State&$\Gamma$&$\alpha$&$\beta$ &$\Gamma_{1}$&$\Gamma_{2}$&$\rm E_{break}$&  $\chi_{\text{red}}^2$ &$\rm F$& $ \rm p-value$ \\
    \hline
    
     S1   & $1.47_{-0.27}^{+0.33}$ &$1.51$ & $-0.20$  &$1.30 $& $2.49$ & $2.19$ & $1.57$, $2.32$, $4.40$ &$0.02, 0.03$& $0.08, 0.09$\\
    
     S2  & $1.10_{-0.23}^{+0.31}$ &$1.48_{-0.20}^{+0.24}$ & $-1.03_{-0.41}^{+0.54}$  & $2.57_{-0.55}^{+0.39}$ & $0.84_{-0.18}^{+0.23}$ & $0.84_{-0.08}^{+0.11}$ &$1.83, 1.70, 1.28$ &$1.38, 2.05$& $0.30, 0.27$ \\
   
     S3  & $1.36_{-0.16}^{+0.19}$  & $1.20_{-0.17}^{+0.21}$ & $0.37_{-0.30}^{+0.41}$ & $0.70_{-0.53}^{+0.43}$ & $1.44_{-0.13}^{+0.17}$ & $0.78_{-0.15}^{+0.17}$ &$ 0.86, 0.91, 1.14$ & $0.76, 0.28 $& $0.42, 0.76$\\
   
    \hline
    \hline
  \end{tabular}
 \caption{Best fit X-ray spectral parameters obtained using power-law (PL), log-parabola (LP) and broken power-law (BPL) models for the states S1, S2 and S3.
Parameters are obtained for the flux unit $erg/cm^2/s$. Due to the high $\chi_{\text{red}}^2$ values for the LP and BPL models in state S1, we were unable to calculate the errors in the parameters.
The F and p-values represents the F-test statistics and the probability of the null hypothesis, respectively.}

\label{tab:x-ray}
\end{table*} 
%%%%%%%%%%%%%%%%%%%%%%%%%%%%%%%%%%%%%%%%%%%%%%%%%%%%

The Swift-Ultra-Violet/Optical Telescope (Swift-UVOT) \citep{2005SSRv..120...95R} has also made the observations for the FSRQ B2\,1348+30B along with Swift-XRT. The observations include all six filters (V: 5468 Å, B : 4392 Å,
U : 3465 Å, W1 : 2600 Å, M2 : 2246 Å, and W2 : 1928 Å) with each filter analysed separately. The photometry was computed for all observations using standard 5 arcsec source region and a background region of approximately 3 times the source region.
Multiple images from the various filters were combined using the "uvotisum" tool.
The magnitudes were determined using the "uvotsource" tool \emph{(HEASOFT -v6.31.1)} and  corrections for extinction were applied \citep{2009ApJ...690..163R}, utilizing the E(B -- V) = 0.0208 reddening coefficient from \citep{2011ApJ...737..103S} data.
%%%%%%%%%%%%%%%%%%%%%%%%%%%%%%%%%%%%%%%%%%%%%%%%%%%%%%%%%%%%%%%%%%%%%%%%%%%%%%%%%%%%%%%%%%%%%%%%%%%%%%%%%%%%%%%%%%%%%%%%%%%%%%%%%%%%%%%%
\begin{figure*}
\centering	
 \includegraphics[width=1\textwidth]{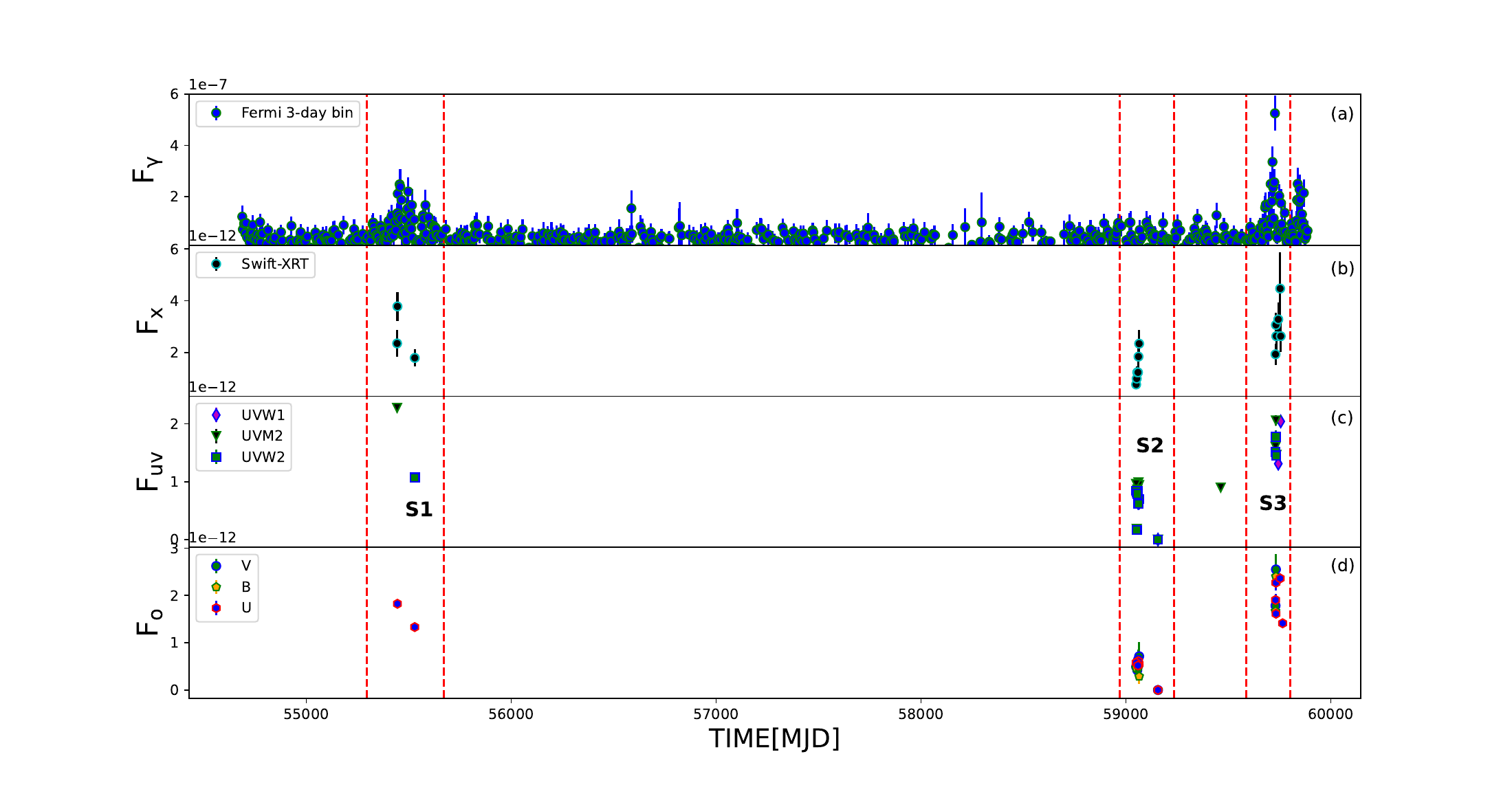}
		%\vspace{0.5cm}
 \caption{The multi-wavelength lightcurve of FSRQ B2\,1348+30B between 2008-08-04 to 2022-11-03 (54682–59886 MJD). (a) 3-day bin $\gamma$-ray lightcurve integrated over the energy range 0.1–300
GeV [(Flux (E$>$100 MeV)] in units of [$\rm ph\,cm^{-2}\,s^{-1}$]. (b) Swift-XRT lightcurve in units of [$ \rm erg\,cm^{-2}\,s^{-1} $] in the 0.3-10 keV energy range. The panels (c) and (d) represent the Swift-UVOT flux points for  W1, M2, W2 V, B and  U bands in units of [$\rm erg\,cm^{-2}\,s^{-1}$]. The time periods S1, S2 and S3 marked by red vertical lines contain the simultaneous observations that are further analysed in this paper.}
		\label{fig:3_day_fermi}
\end{figure*}

\section{RESULTS AND DISCUSSION}
\label{sec:results}
\subsection{Temporal Behaviour}
We analysed the \emph{Fermi}-LAT data  of the FSRQ B2\,1348+30B for the selected duration and the 3-days binned $\gamma$-ray lightcurve, integrated 
over the energy range 0.1--300\,GeV, is shown in Fig.\,\ref{fig:3_day_fermi} (top panel). 
%for the period spanning from 2008-08-04 (54682 MJD) to 2022-11-03 (59886 MJD). 
In the bottom panels of Fig.\,\ref{fig:3_day_fermi}, we show the X-ray, UV and optical  lightcurves. Simultaneous observations
at all these energy bands were available only for three epochs 2010-04-10 to 2011-04-19 (55296--55670 MJD), 2020-05-01 to 2021-01-22 (58970--59236 MJD) and
 2022-01-09 to 2022-08-11  (59588--59802 MJD) which are marked as 
S1, S2 and S3. The average $\gamma$-ray flux corresponding to these epochs are $\rm (5.3\pm3.2)\times10^{-8}\,ph\,cm^{-2}\,s^{-1}$, 
$\rm (2.0\pm1.7)\times10^{-8}\,ph\,cm^{-2}\,s^{-1}$ and $\rm (7.5\pm3.4)\times10^{-8}\,ph\,cm^{-2}\,s^{-1}$, respectively.

\begin{figure*}
    \centering
\includegraphics[scale=0.45]{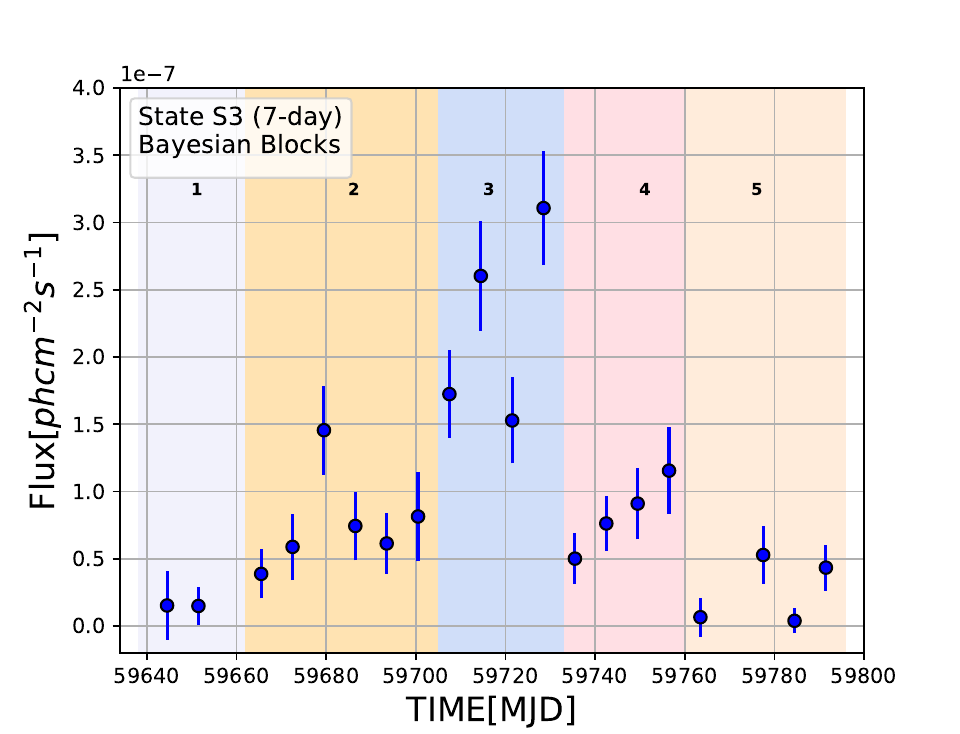}
  %  \hspace{0.5cm}
    \includegraphics[scale=0.45]{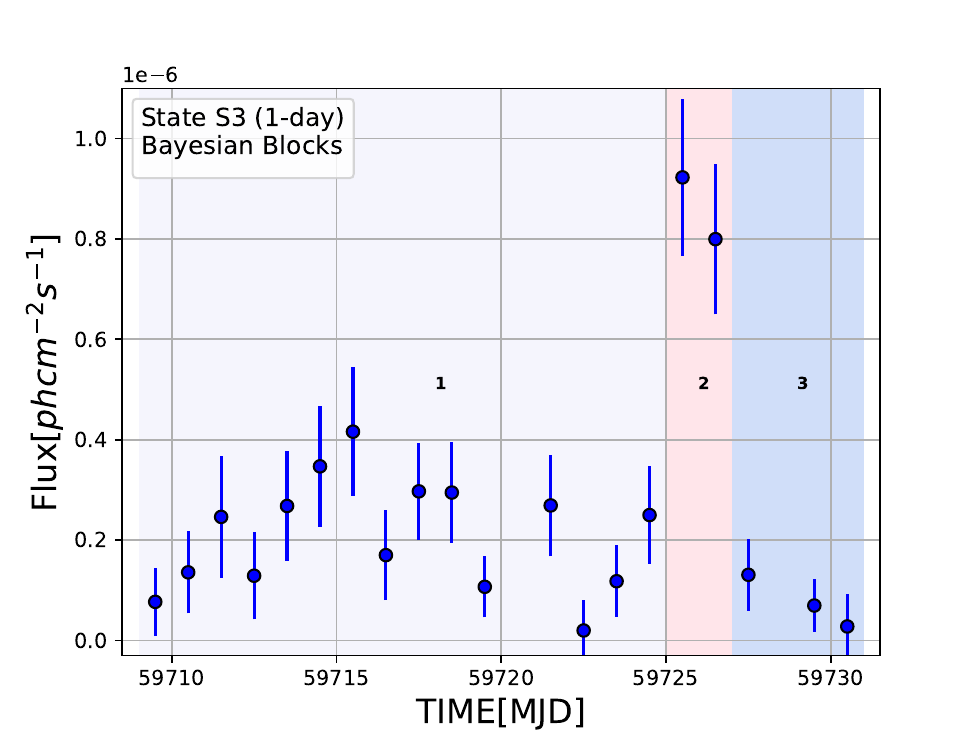}
    \caption{Lightcurve of the state S3 in 7-day (left) and 1-day (right) bining. Bayesian analysis approach conducted for defining solid blocks and estimating variability timescale.}
    \label{fig:baysn_grph}
\end{figure*}

The source showed significant flux variability, especially in the  $\gamma$-ray band with highest flaring activity observed during epochs S1 and S3.
During the epoch S2 the source was relatively in the low $\gamma$-ray activity state.  
The two flaring epochs, S1 and S3 were further investigated with different time binning. Initially, weekly binned lightcurves are generated for each flaring 
period to detect variations in $\gamma$-ray flux. Further, the Bayesian blocks method \citep{2013ApJ...764..167S} is employed (\emph{astropy} ver. 4.2) 
for flare S3 to identify the flux variations within optimised time intervals while assuming a false alarm probability of $0.05$. This study revealed 
five blocks for the flare with the third block as the most brightest one. The Bayesian blocks along with the weekly averaged lightcurve is shown in 
Fig.\ref{fig:baysn_grph} (left). The duration corresponding to block 3 was again investigated using 1-day averaged bins and the Bayesian analysis showed 3 subblocks suggesting 
significant flux variation. The subblock 2 was found to be the shortest and lasts for $\sim$ 3\,days and shown in Fig.\ref{fig:baysn_grph} (right).

The variability timescale $\tau$ of FSRQ B2\,1348+30B, obtained as $\leq$ 3\,days, can be used to provide  
estimates on the emission region
size and its location from the central black hole. From the light travel time arguments, the size of the emission region can be limited to 
$R \leq c \delta \tau/(1+z)$ where, $\delta$ is the relativistic Doppler factor of the blazar jet. From the obtained $\tau$ we find the emission region size
to be $R \leq 9\times10^{16}\,cm$ for the choice of $\delta\approx 20$.
The distance from the central black hole estimated from the travel time 
during $\tau$ will be $d_{\gamma} \sim 2\,c\,\tau \delta^2/(1+z)$ and this is found to be $3.7 \times 10^{16} \, \text{cm}$.

%%%%%%%%%%%%%%%%%%%%%%%%%%%%%%%%%%%%%%%%%%%%%%%%%%%%%%%%%%%%%%%%%%%%%
%%%%%%%%%%%%%%%%%%%%%%%%%%%%%%%%%%%%%%%%%%%%%%%%%%%%%%%%%%%%%%%%%%%%%%

\begin{figure}
		\centering	%\includegraphics[width=1\textwidth,height=0.5\textheight]{fit_lc.pdf}
  \includegraphics[width=0.45\textwidth]{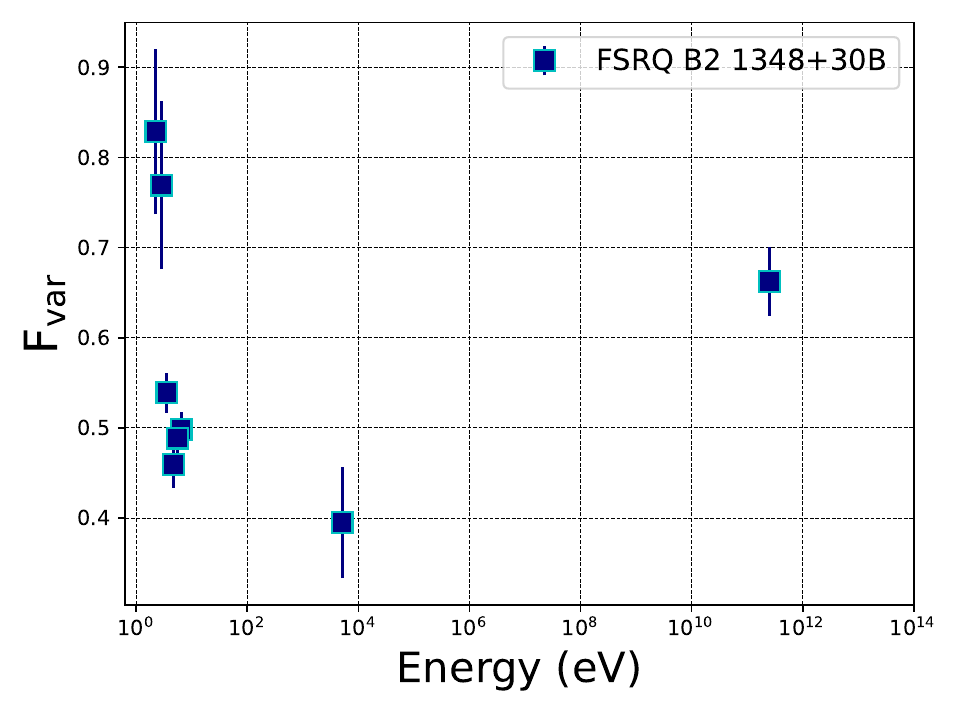}
		%\vspace{0.5cm}
 \caption{Fractional variability amplitude ($F_{var}$) obtained in various energy bands is plotted against the energy  during 2008-08-04 to 2022-11-03 (54682–59886 MJD).}
		\label{fig:frac_varg}
\end{figure}

The continuous monitoring of the source using \emph{Fermi} lets us to quantify the variability behaviour of the source in $\gamma$-ray energy band.
However, the source lacks continuous observations in X-ray and Optical/UV bands and hence, the flux variability estimates at these energies are 
limited by the available data set.  Using the lightcurve with $N$ flux points, the fractional RMS variability amplitude can be 
estimated as \citep{2003MNRAS.345.1271V}
\begin{equation}
\centering 
    F_{var} = \sqrt{\frac{S^{2} - \langle \sigma_{err}^{2} \rangle}{{\langle f \rangle}^{2}}}
    \label{eq:var_eqn}
\end{equation}
where, $\langle f \rangle$ is the mean flux, $S^2$ is the variance and the mean square error given by   
\begin{equation}
\centering
    \langle \sigma^2_{\text{err}} \rangle = \frac{1}{N} \sum_{i=1}^{N} \sigma^2_{\text{err},i}
\end{equation}
with $\sigma_{\text{err},i}$ being the error in individual data points.
The uncertainty in $F_{var}$ can be obtained from the relation \citep{2003MNRAS.345.1271V}
\begin{equation}
    \label{eq:_eqn}
    (F_{\text{var}})_{\text{err}} = \sqrt{\frac{1}{2N} \left(\frac{\langle \sigma_{\text{err}}^{2} \rangle}{F_{\text{var}} \langle f \rangle^{2}}\right)^2 + \frac{1}{N}\frac{\langle \sigma_{\text{err}}^{2} \rangle}{\langle f \rangle^{2}}}
\end{equation}
In Table\,\ref{tab:frac_vart}, we provide the estimated $F_{var}$ for the different energy bands along with the uncertainties and in 
Fig.\ref{fig:frac_varg}, we plot $F_{var}$ against the photon energy. We find the source is significantly variable in optical
and $\gamma$-ray energies with the variability amplitude greater than 50\% while it is moderately variable in the 
UV and X-ray energy band ($<50\%$). This plausibly indicates that the emission at $\gamma$-ray and the optical 
energy bands may be associated with the high energy electrons while low energy electrons may be responsible
for the X-ray emission. A similar result was also obtained from the broad-band spectral modelling of the source
using synchrotron and inverse Compton emission processes (see \S\,\ref{subsec:brd_mdl}). Since the energy loss rate of the emitting
electrons under these emission processes depends on the square of the particle energy, the emission arising from the 
high energy electrons is expected to be more variable than the low energy ones. Interestingly, the UV band appears
to be less variable though broad-band SED modelling suggests that this emission to be associated with the high energy
electrons. A possible reason may be the contribution of significant non-variable accretion disk emission along with 
the variable jet emission and this cumulatively reduces the effective variability.

%%%%%%%%%%%%%%%%%%%%%%%%%%%%%%%%%%%%%%%%%%%%%%%%%%%%%%%%%%%%%%%%%%%%%%%%%%%%%%%%%%%%
%%%%%%%%%%%%%%%%%%%%%%%%%%%%%%%%%%%%%%%%%%%%%%%%%%%%%%%%%%%%%%%%%%%%%%%%%%%%%%%%%
\begin{table}
\centering
\renewcommand{\arraystretch}{1.2} 
\begin{tabular}{l c}
\hline 
\hline
Energy band  & $\rm F_{var}$ \\
\hline

$\gamma$-ray (\mbox{0.1--300 GeV}) & $0.66 \pm 0.03$\\
X-ray (0.3--10 keV) & $0.39 \pm 0.06$\\
UVOT band-W2 & $0.32 \pm 0.05$ \\
UVOT band-M2 & $0.48 \pm 0.02$\\
UVOT band-W1 & $0.45 \pm 0.02$\\
UVOT band-U &$ 0.53 \pm 0.02$\\
UVOT band-B & $0.82 \pm 0.09$\\
UVOT band-V & $0.76 \pm 0.09$\\
\hline
\hline
\end{tabular}
%\vspace{0.5cm}
 \caption{Fractional variability amplitude obtained in optical, UV, X-ray and $\gamma$-ray energy bands plotted as a function of their energy during 2008-08-04 to 2022-11-03 (54682–59886 MJD).}
\label{tab:frac_vart}
\end{table}

%%%%%%%%%%%%%%%%%%%%%%%%%%%%%%%%%%%%%%%%%%%%%%%%%%%%%%%%
\subsection{Spectral Behaviour}
\label{sec:spec_behv}
\begin{figure*}
    \centering
\includegraphics[scale=0.45]{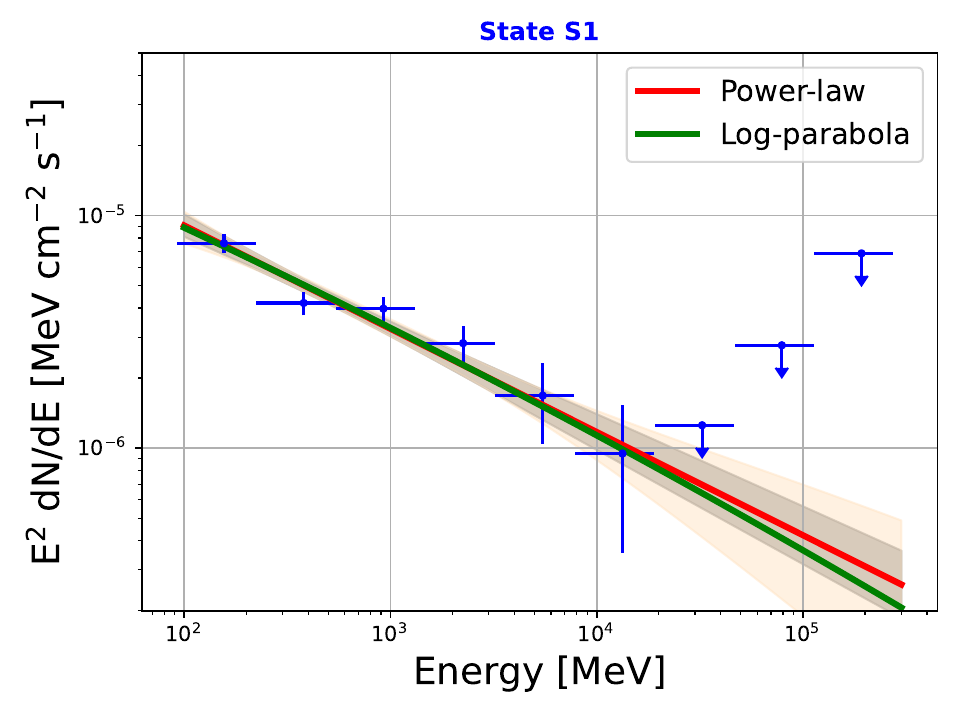}
    \hspace{0.5cm}
 \includegraphics[scale=0.45]{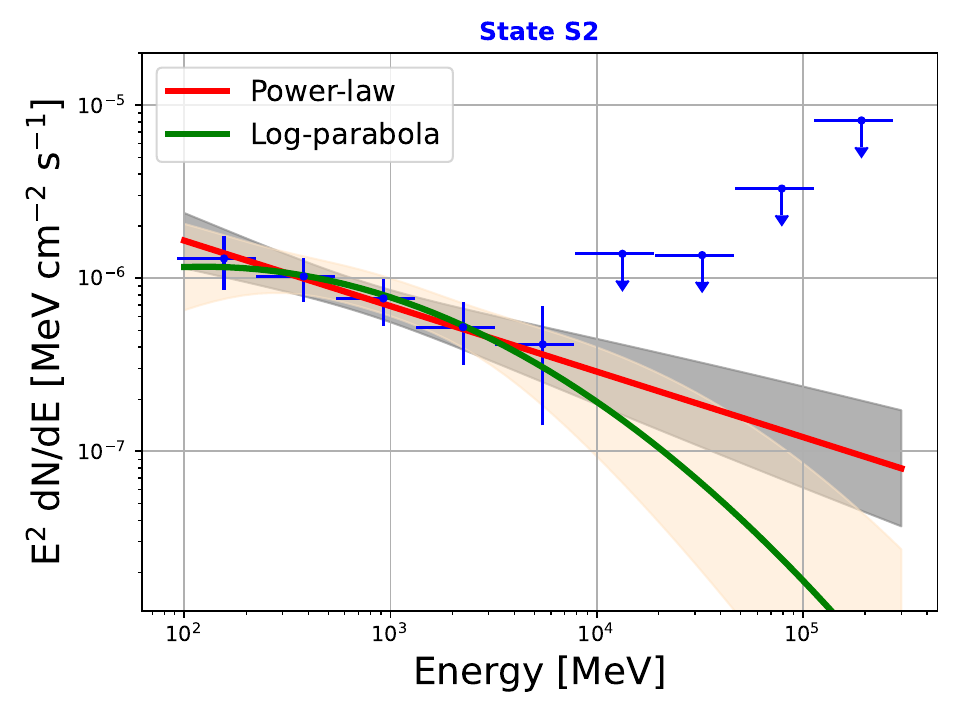}
 \includegraphics[scale=0.45]{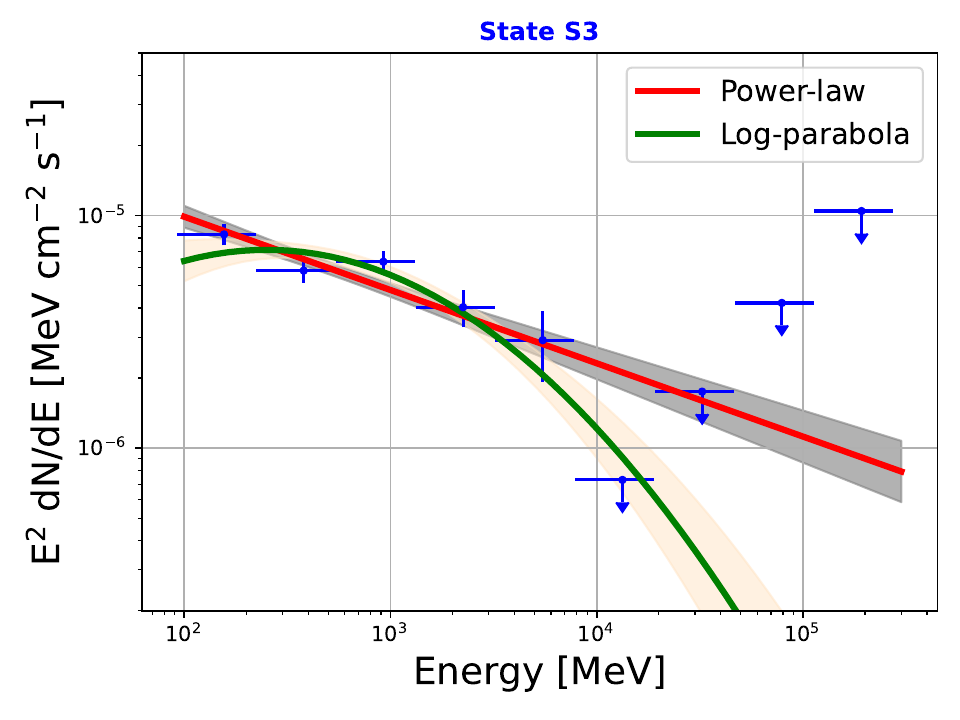}
     \hspace{0.5cm}
 \includegraphics[scale=0.45]{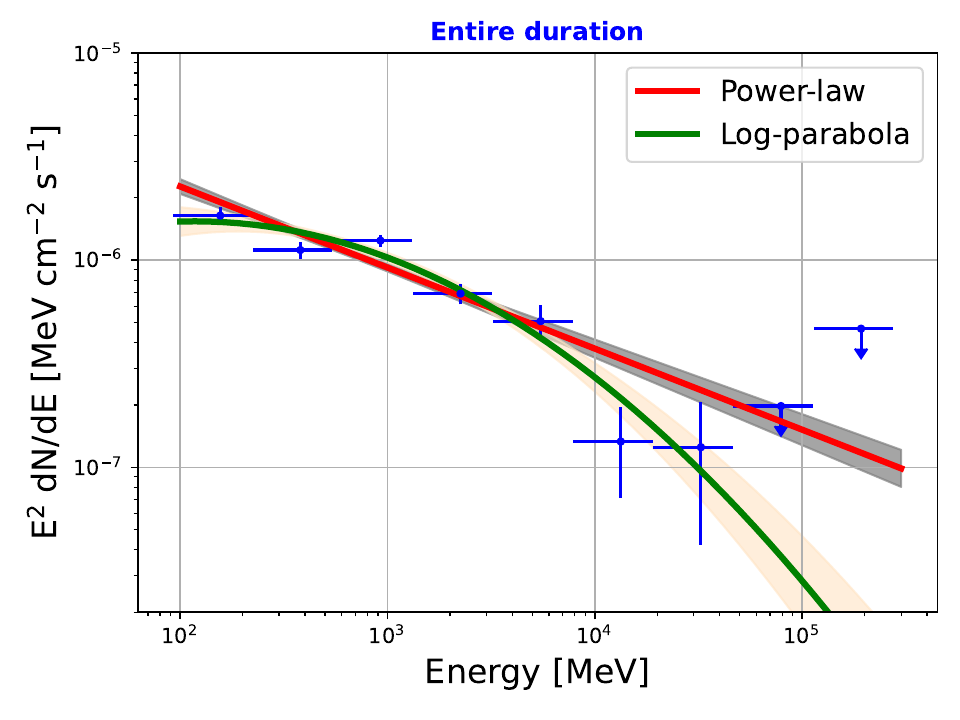}
    \caption{Spectral fit to $\gamma$-ray SEDs during the states S1, S2 and S3 having simultaneous multi-wavelength observations, and for the entire $\gamma$-ray observation duration. The spectral fit models are power-law (red line) and log-parabola (green line). } 
%Top left corresponds to state S1, top right to state S2, bottom left to state S3 and bottom right shows the the average spectrum for the entire duration.

    \label{fig:spect_gray}
\end{figure*}

The $\gamma$-ray spectrum of the FSRQ B2\,1348+30B during the epochs S1, S2 and S3 are studied in the energy range from 
0.1--300\,GeV. 
%The study analysed the spectra during both quiescent and active phases individually, based on the results of the $\gamma$-ray lightcurve. 
The SED points for each state are acquired by dividing the total energy range of 0.1 – 300 GeV into 10 equal logarithmically spaced bins.
The energy resolved bins are fitted using log likelihood analysis. During the spectral fit, the parameters for all  sources (other than B2\,1348+30B) 
falling within the ROI are 
frozen to their best-fit values.
Two specific spectral models were used to 
characterize the $\gamma$-ray spectral behaviour, the power-law model (PL)
\begin{equation}
\label{eqn:pwr_lw}
\frac{dN}{dE} = N_0 \left(\frac{E}{E_0}\right)^{-\Gamma}\\
\end{equation}
where, $N_{0}$ is the spectral normalization at the scale energy $E_{0}$ with $\Gamma$ being the PL index, and the log-parabola model (LP) 
\begin{equation}
\frac{dN}{dE} = N_0 \left(\frac{E}{E_0}\right)^{-(\alpha +\beta\,log(E/E_0))}
    \end{equation}
where, $\alpha$ is the spectral slope at the scale energy $E_{0}$ and $\beta$ is the parameter deciding the peak curvature. The spectral fitting was 
performed for the flux states S1, S2 and S3, and also for the average $\gamma$-ray spectrum obtained from the entire duration considered here.
These spectral fits along
with the observed data are shown in Fig.\,\ref{fig:spect_gray} and best fit parameters are given in Table\,\ref{tab:spect_table}. To identify whether the $\gamma$-ray
spectra showed statistically significant curvature, we estimated the test statistics $\rm TS_{curve} = 2\,[log \mathcal{L}(LP) - log \mathcal{L}(PL)]$, 
with $\mathcal{L}$ being the likelihood function, for all the states \citep{2012ApJS..199...31N}. The curvature can be asserted if $\rm TS_{curve} \geq 16$;
however, this criteria is not satisfied for any of the spectral fit (Table\,\ref{tab:spect_table}, last column). Hence, we conclude the $\gamma$-ray spectra is better represented by 
a power-law function. We also extended the analysis to identify the maximum energy of the $\gamma$-ray photon emitted within the entire duration of observation
considered here. We did not observe photons with energy $\gtrsim$ 20\,GeV and significance exceeding 3$\sigma$ for this source.

%%%%%%%%%%%%%%%%%%%%%%%%%%%%%%%%%%%%%%%%%%%%%%%%%%%%%%%%%%%%%%%%%%%%%%%%%%%%%%
%%%%%%%%%%%%%%%%%%%%%%%%%%%%%%%%%%%%%%%%%%%%%%%%%%%%%%%%%%%%%%%%%%%%%%%%%%%%%%
\begin{table*}
\centering
\renewcommand{\arraystretch}{1.3} 
      \begin{tabular}{lccccccc} \\
\hline
\hline
    Period & Time & Model & Flux & $\Gamma$ & $\alpha$ & $\beta $ & $\rm TS_{curve}$ \\
    & (MJD) & &($10^{-8}\rm ph\,cm^{2}s^{-1}$)  \\
       \hline
       
      State S1 &55296-55670&PL & $6.28\pm0.05$ & $2.44\pm0.05$ & -- & -- \\
      &&LP&$6.23\pm0.04$&& $2.45\pm 0.05$  &$0.006\pm--$& 0.03 \\
      State S2 &58970-59236&PL&  $1.19\pm0.03$ & $2.37\pm0.01$ & -- & -- \\
      &&LP&$1.04\pm0.03$&& $2.31\pm 0.22$ & $0.09\pm--$ &0.44   \\
      State S3 &59588-59802&PL&  $7.52\pm0.06$ & $2.31\pm0.05$ & -- & --  \\
      &&LP&$6.47\pm 0.04$&& $2.29\pm 0.06$  &$0.12\pm0.04$ & 8.94 \\
      Entire duration &54690-59890 &PL&$1.63\pm 0.01$ & $2.39\pm0.03$ & -- & -- \\
      &&LP&$1.38\pm0.02$&&$2.33\pm0.04$ &$0.08\pm0.02$& 10.2 \\
      \hline
    \hline   
\end{tabular} 
  %\vspace{0.5cm}
 \caption{ The spectral analysis results of the $\gamma$-ray data during various flux states.} 
    \label{tab:spect_table} 
   \end{table*}
%%%%%%%%%%%%%%%%%%%%%%%%%%%%%%%%%%%%%%%%%%%%%%%%%%%%%%%%%%%%%%
%%%%%%%%%%%%%%%%%%%%%%%%%%%%%%%%%%%%%%%%%%%%%%%%%%%%%%%%%%%%%%

The X-ray spectra of the source was well represented by a power-law during the high flux states S1 and S3 while a broken power-law represents the spectrum
during the low flux state S2 (Based on the $\chi^2_{\rm red}$ provided in Table\,\ref{tab:x-ray}). 
However, we also applied F-test \citep{2003drea.book.....B} to compare power-law with log-parabola and broken power-law models. The resulting p-values were > 0.05 supporting power-law  as the best  fit model across all the three states.
Further, the X-ray spectra are harder than the $\gamma$-ray spectra for the 
states S1 and S3 and this suggests the former fall on the low energy end of the Compton spectral component. Hence, low energy electrons will be responsible
for the X-ray emission compared to the $\gamma$-ray emission. This is consistent with the conclusions arrived from the temporal analysis where the $\gamma$-ray 
emission is found to be more variable than the X-ray emission. Interestingly, for the quiescent state S2 the broken power-law spectral fit suggests the high energy 
index is harder than the low energy index. Such scenario supports that the X-ray spectrum falls on the transition region between the synchrotron and 
Compton spectral components. Nevertheless, the extreme hard high energy index ($\sim 0.8$) cannot be attained through IC process since it demands a particle
distribution with index $\sim 0.6$ and this cannot be easily attained through shock acceleration \citep{2007Ap&SS.309..119R}. Alternatively, such hard 
X-ray spectrum can also be attributed to the low energy cut-off of the underlying electron distribution and can provide constraints on the kinetic power of the 
blazar jet (\S\,\ref{subsec:jet_kp}).

To identify whether the UV/optical data available from \emph{Swift}-UVOT had the presence of emission components additional 
to the jet emission, we fitted the observed fluxes with a power-law. For the states S2 and S3, the flux information was available from all the six filters; 
whereas, fluxes from only three filters were available during the state S1. Initial fit using a power-law resulted in poor statistics 
($\chi^2_{\rm red} \gtrsim 10$). We repeated the fitting by adding arbitrary error to the data since this will give an estimation about the deviation of 
the optical/UV spectra from a power-law. Our study suggested that in order to obtain a acceptable statistics one needs to add $\gtrsim$ 10\% of error and 
this suggests significant presence of other emission components additional to the jet emission in the optical/UV spectra.
The additional emission can likely be associated with the accretion disk, or potentially other thermal contributions in the optical/UV regime.
 This result was consistent with the temporal study where minimal variability in UV band supports the presence of additional emission components.

\subsection{Broad-band SED Modelling}
\label{subsec:brd_mdl}
Availability of simultaneous data during the epochs S1, S2 and S3 lets us to perform a broad-band spectral modelling of the 
source using synchrotron, SSC and EC processes. For this we used the one-zone leptonic model described in \citep{2018RAA....18...35S}.
The model assumes the emission region to be a spherical blob of non-thermal electron distribution moving down the jet relativistically.
The electron distribution is assumed to be a broken power-law described by
\begin{equation}
N(\gamma)d\gamma = 
\begin{cases}
K \gamma^{-p}d\gamma & \text{for } \gamma_{\min} <\gamma< \gamma_b, \\
K \gamma_b^{q-p}\gamma^{-q}{\rm exp}(-\gamma/\gamma_{\rm max})d\gamma& \text{for } \gamma> \gamma_b
\end{cases}
\end{equation}
where, $K$ is the normalization factor, $p$ and $q$ are the high and low-energy indices, $\gamma_b$ is the electron Lorentz factor 
corresponding to the break in the particle distribution 
and $\gamma_{\rm max}$ decides the high energy cut-off. We represent the magnetic field as B and the size of the emission region as R.
The electron distribution mainly radiate through synchrotron, SSC and EC/IR processes. The external photon field for the EC/IR process
is chosen to be a blackbody at temperature 1000 K. Due to the relativistic motion of the emission region, the rest frame emission will 
be boosted by the Doppler factor $\delta = 1/\Gamma(1 - \beta \cos \theta)$ where, $\Gamma$ is the bulk Lorentz factor of the jet, $\theta$
is the viewing angle and $\beta$ is the dimensionless jet velocity. We also impose equipartition between the magnetic field and the electron
energy densities as an additional constraint. The emissivities due to these radiative processes are evaluated numerically and the code was 
coupled as a local model in the XSPEC to perform the broad-band spectral fitting. The fitting was first performed with typical parameters as
initial guess. However, due to limited information available the confidence intervals were obtained only for $p$, $q$, $\gamma_{min}$ and  $B$ while rest of the parameters are frozen to its best fit value. The injection energy of the electron distribution $\gamma_{\rm min}$ is constrained from the 
extremely hard X-ray spectrum corresponding to state S2 (see, \S\,\ref{sec:spec_behv}). In Table \ref{tab:Brd_SED_parms}, we provide the best 
fit parameters for each state and the model curves along with the observed fluxes are shown in Fig.\,\ref{fig:bb_sed}.
Though the model represents the observation reasonably well, still the values of the $\chi^2$ provided in the Table \ref{tab:Brd_SED_parms}
are large. Such large $\chi^2$ are mainly introduced by the optical/UV fluxes where the uncertainties are very small. We investigated this
further by estimating the $\chi^2$ after excluding the optical/UV data but for the same set of model parameters. We obtained the $\chi^2$/dof values as 
11/4 for S1, 19/6 for S2 and 18/7 for S3, which are significantly smaller than the ones quoted in Table \ref{tab:Brd_SED_parms}. As mentioned in the earlier sections, 
there could be additional emission contribution at
this energy band and simple emission model involving synchrotron and inverse Compton processes alone may not be sufficient enough to explain 
the optical/UV emission.
%\textbf{However, the value of the $ \chi^2/\mathrm{dof}$ is high, which is likely due to certain flux points deviating from the model. These deviations may arise from systematic uncertainties in the data or could indicate the presence of a more complex emission mechanism that may not be accounted for by the simple model employed.}

From the fit we find the low energy particle index is relatively hard during the flaring states. Since the particle 
index is governed by the rate of acceleration \citep{1998A&A...333..452K}, this result suggests efficient particle 
acceleration to happen during the flaring states (plausibly initiated by a strong shock). The magnetic field on the other 
hand, is large during the quiescent  state compared to the flaring states. The development of shock can transfer the bulk energy 
to the particle distribution and the magnetic field. Hence, this result is contrary to the persumption that the flaring states 
are triggered by strong shocks. A plausible scenario can be, the particle acceleration and most of the emission can happen at two
physically seperated regions (two-zones) with the physical dimension of the latter much smaller. The striking result obtained through 
the broad-band SED fitting is that the significant increase in the bulk Lorentz factor (nearly two times) during the flaring 
states S1 and S3 compared to the quiescent state S2. This enhanced bulk energy of the jet can allow significant fraction of it to be 
transferred to the particle distribution resulting in a flare. Hence, this study supports that the flaring episode of the blazar 
may be associated with strong shocks initiated by the high speed ejecta from the central black hole.

%%%%%%%%%%%%%%%%%%%%%%%%%%%%%%%%%%%%%%%%%%%%%%%%%%%%%%%%%%%%%
\subsection{Jet Kinetic Power}
\label{subsec:jet_kp}
\begin{figure*}
    \centering
\includegraphics[width=0.3\textwidth,height=0.3\textheight,angle=270]{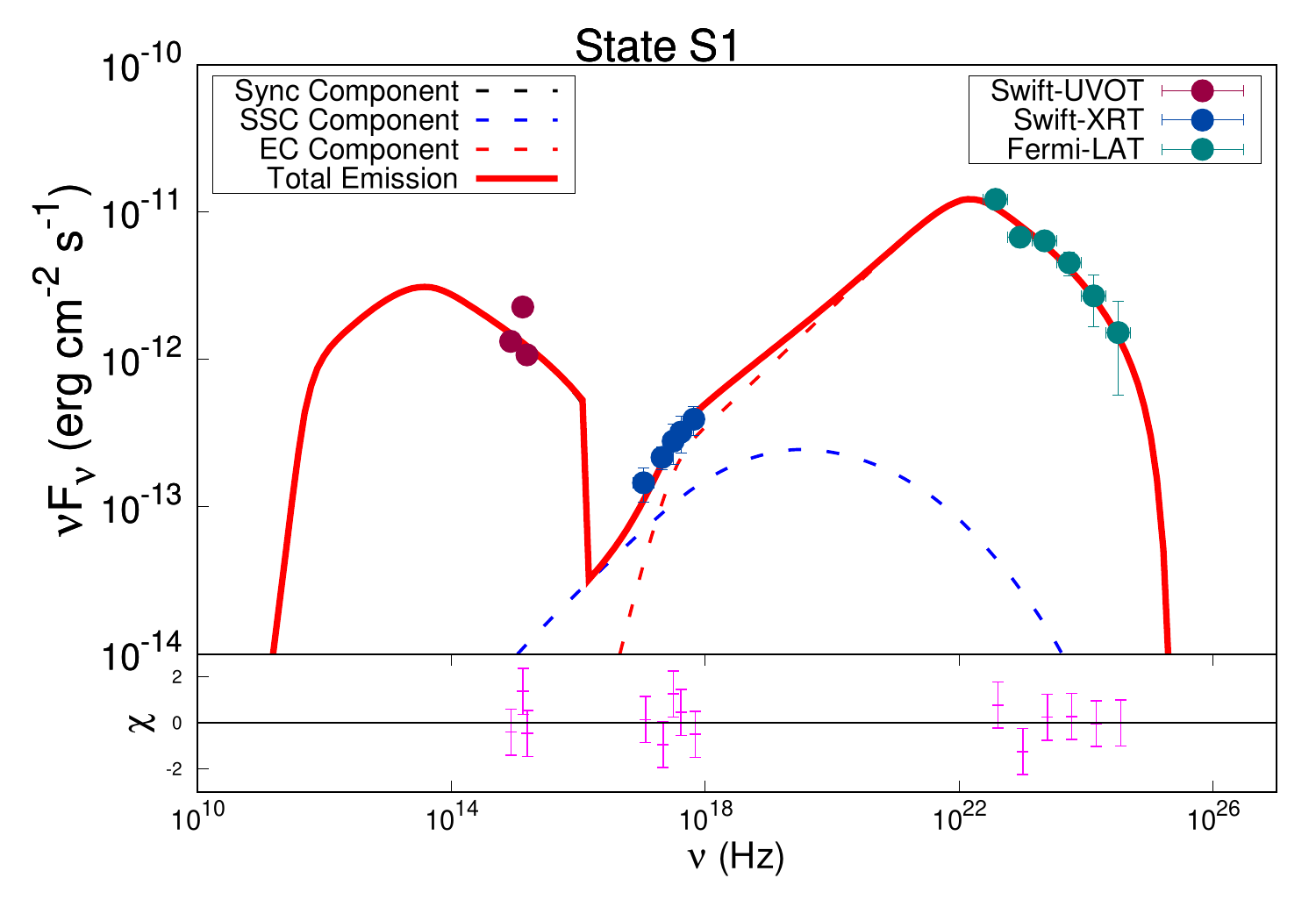}
 %  \hspace{0.5cm}
 %\vspace{0.4cm}
 \hspace{0.4cm}
\includegraphics[width=0.3\textwidth,height=0.3\textheight,angle=270]{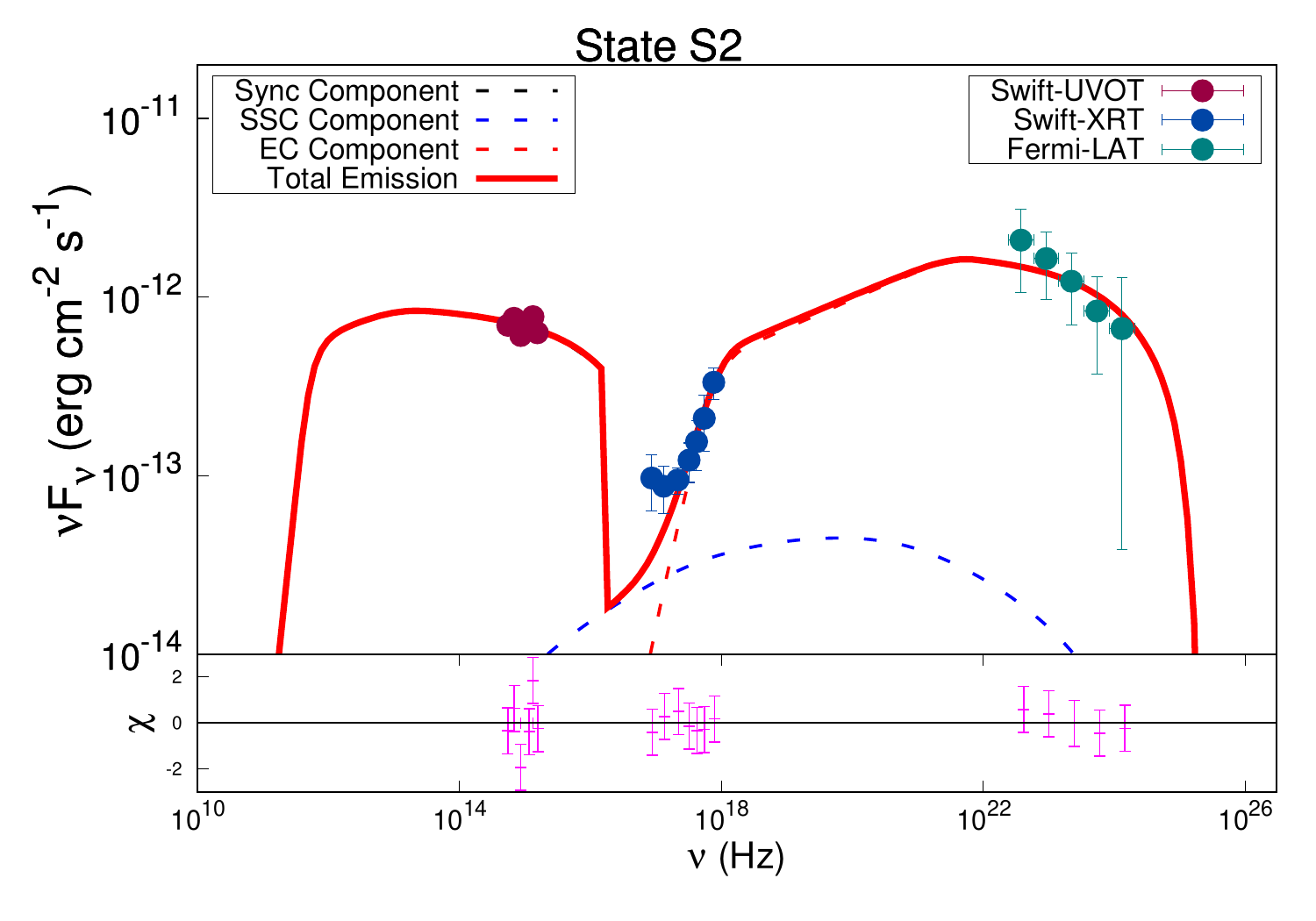}
%\vspace{0.4cm}  
\includegraphics[width=0.3\textwidth,height=0.3\textheight,angle=270]{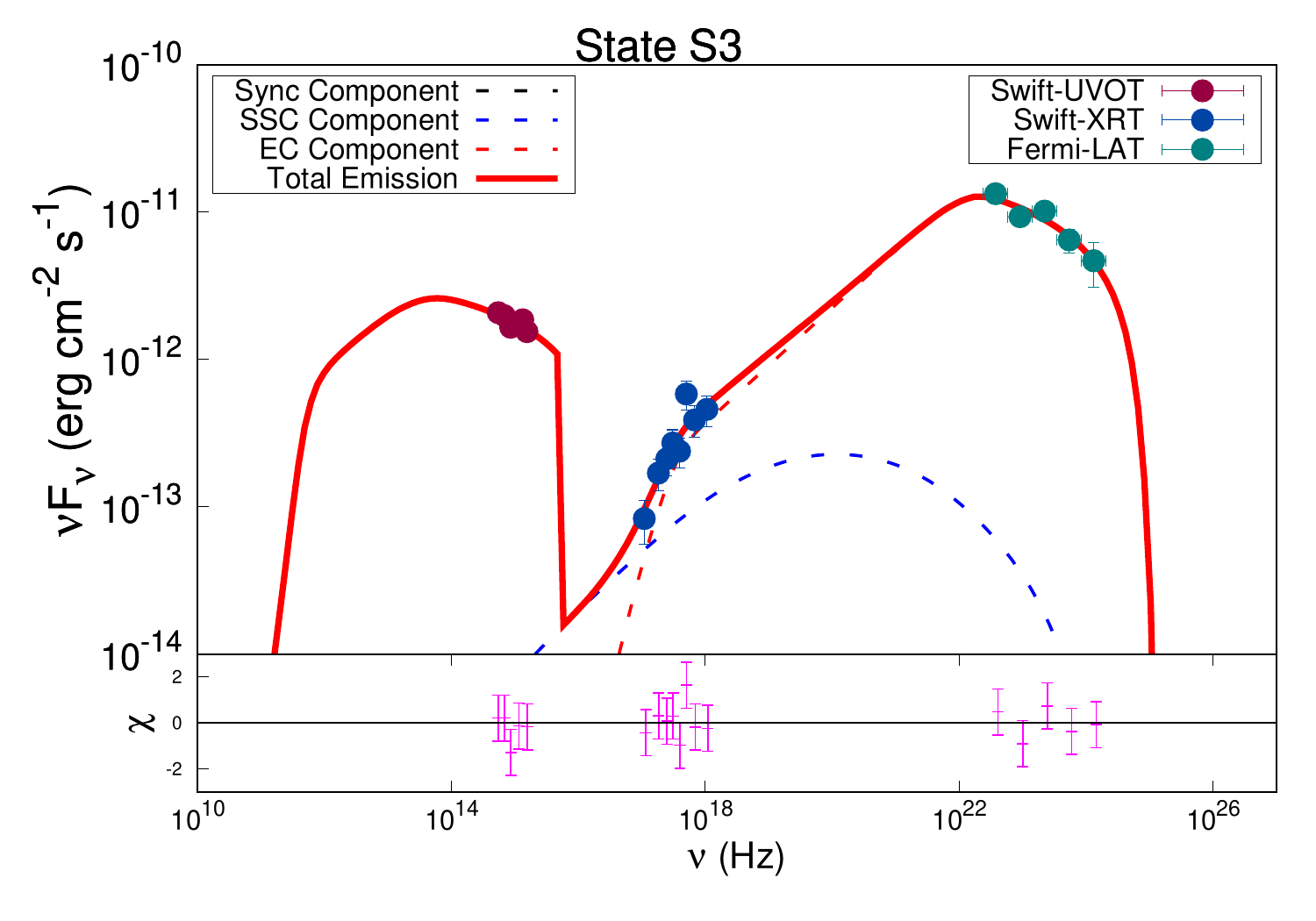}
    \caption{Broad-band SED plots of the FSRQ B2\,1348+30B, during ﬂaring states S1, S3 and quiescent state S2. The observed flux points and the corresponding emission components are depicted in legends of each figure, respectively.}
    \label{fig:bb_sed}
\end{figure*}

\begin{figure*}
    \centering
\includegraphics[scale=0.5]{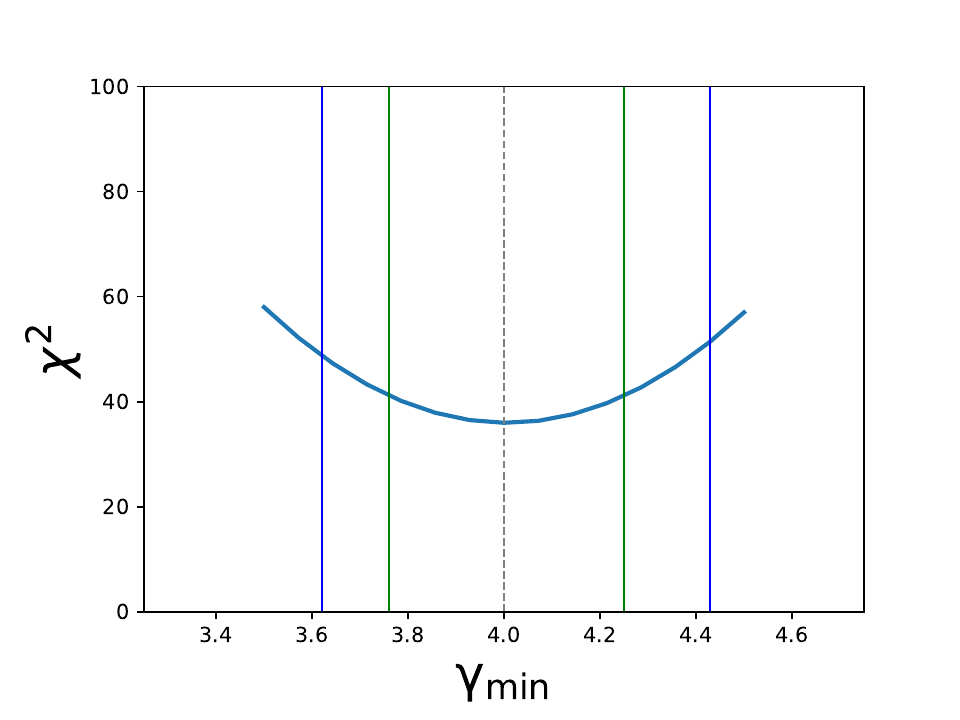}
  %  \hspace{0.5cm}
    \includegraphics[scale=0.5]{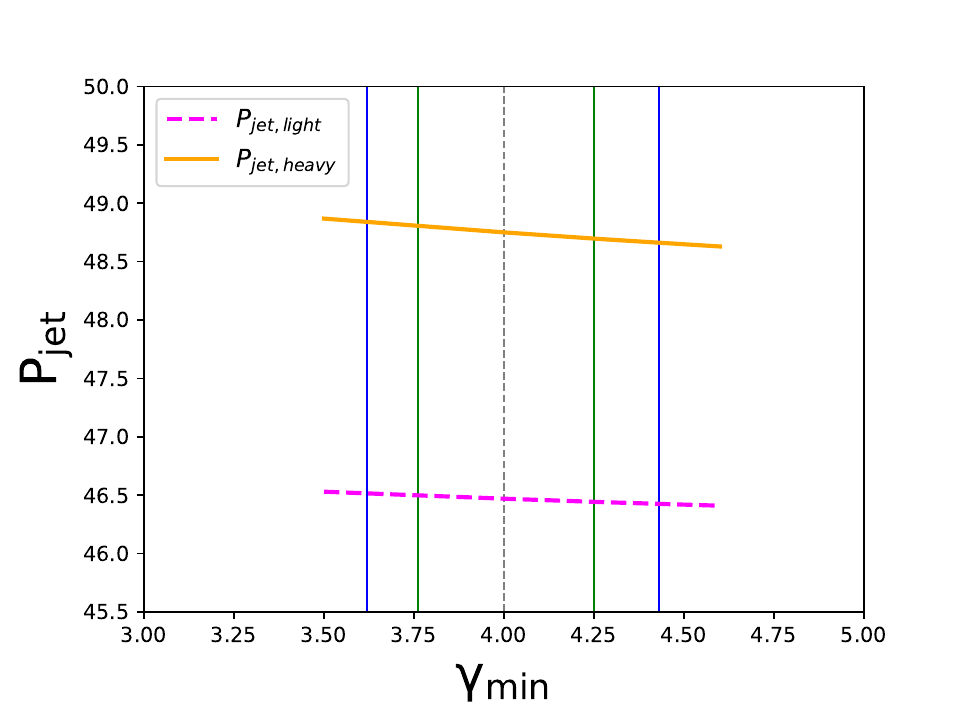}
    \caption{Left plot: Solid dodgerblue line denotes the variation of $\chi^{2}$ with minimum electron energy ($\gamma_{min}$). Right plot: Variation of jet power as function of $\gamma_{min}$. For both the plots: The dashed grey lines represent the best fit value of $\gamma_{min}$. The green and blue vertical lines correspond to 1$\sigma$ and 2$\sigma$ uncertainties on $\gamma_{min}$, respectively.}
    \label{fig:pjet_var}
\end{figure*}

The extremely hard X-ray spectrum of the state S2 led us to constrain the $\gamma_{\rm min}$ of the electron distribution. From the observed 
SED (Fig.\,\ref{fig:bb_sed}), the minimum emitted EC/IR photon frequency will be $\nu_{\rm min}\sim$ $2.26\times10^{17}$Hz. This can be translated to the 
minimum energy of the non-thermal electron distribution in terms of the other jet parameters as \citep{2018RAA....18...35S}
\begin{equation}
\gamma_{\text{min}} = \left( \frac{1+z}{\delta\, \Gamma} \cdot \frac{\nu_{\text{min}}}{\bar{\nu}} \right)^{1/2}
\end{equation}
where, 
$\bar{\nu}= 5.86\times 10^{13}\,(T/1000K)$ Hz. 
To estimate $\gamma_{\rm min}$, we fit the broad-band SED of the state S2 with different choices of $\gamma_{\rm min}$.
In Fig.\,\ref{fig:pjet_var} (left), we show the variation in $\chi^2$ and it is evident that the best fit is obtained when 
$\gamma_{\rm min}= 4$.
The 1$\sigma$ and 2$\sigma$ confidence interval on this estimate is shown as vertical lines in Fig.\,\ref{fig:pjet_var}.

For a power-law electron
distribution (Equation\,\ref{eqn:pwr_lw}), the energy integrated number density of the non-thermal electron distribution will be
\begin{align}\label{eq:eden}
	N_{\rm int} &= \int\limits_{\gamma_{\rm min}}^{\gamma_{\rm max}} \,N(\gamma) d\gamma \nonumber \\
	&\approx \frac{K}{p-1} \gamma_{\rm min}^{-p+1} \quad {\rm for}\; \gamma_{\rm max},\gamma_p \gg \gamma_{\rm min}\; {\rm and }\; p>1,
\end{align}
and the energy density
\begin{align}\label{eq:ue}
	U_e &= m_e c^2 \int\limits_{\gamma_{\rm min}}^{\gamma_{\rm max}} \gamma\,N(\gamma) d\gamma \nonumber \\
	&\approx \frac{K m_e c^2}{p-2} \gamma_{\rm min}^{-p+2} \quad {\rm for}\; \gamma_{\rm max},\gamma_p \gg \gamma_{\rm min}\; {\rm and }\; p>2.
\end{align}
If the matter content of the jet is dominated by electron-positron pairs (light jet) then the kinetic power 
of the jet will be \citep{2008MNRAS.385..283C}
\begin{equation}
    P_{\text{jet,light}} = \pi R^2 \Gamma^2c\,U_e 
\end{equation}
From the best fit parameters, we find the jet power in this 
case to be $2.57\times 10^{46}$ erg/s and this sets the lower limit 
of the blazar jet power.

Considering the jet matter to be purely electron-proton plasma can provide the upper limit on the jet power. For simplicity,
we assume the protons to be cold and do not participate in the radiative processes and additionally, their number is 
equal to that of non-thermal electrons. The jet power will then be 
\begin{align}
    P_{\text{jet,heavy}} &= \pi R^2 \Gamma^2c\,(U_p + U_e)\nonumber \\
    &\approx \pi R^2 \Gamma^2c\,U_p \quad {\rm for} \;U_e \ll U_p
\end{align}
where, $U_p = N_{\rm int} m_p c^2$ is the proton energy density. The condition $U_p<<U_e$ is valid 
until $\gamma_{\rm min}\ll m_p/m_2$ \citep{2021MNRAS.506.3996J}. From the best fit parameters, we find the jet power estimated under this
scenario to be $4.17\times 10^{48}$ erg/s. If we relax the cold proton condition, when significant hadronic emission component is present in the broad-band
SED, then the estimated $P_{\text{jet,heavy}}$ may be higher than the one estimated here. In Fig.\,\ref{fig:pjet_var} (right), we show the blazar jet power
constrained between these limits and confidence intervals of the $\gamma_{\rm min}$. 

The Eddington Luminosity corresponding to the blackhole mass of $1.8\times 10^8\,M_\odot$ for the B2\,1348+30B is $2.34\times10^{46}$ erg/s.
Interestingly, this power is similar to the one estimated for the light jet but nearly two orders less than that of the heavy jet. This may 
raise a question that whether the AGN jets are powered only by the accretion process or additional energy are extracted probably from the 
spin of the blackhole \citep{2014Natur.515..376G}.
In both scenarios, the magnetic field plays a key role in transferring energy from the black hole or accretion disk to the jet \citep{2018ApJ...852..112K}. 
Since the magnetic field is believed to be sustained by accretion, it is expected that the jet power is correlated with accretion power. 
One plausible way to extract the energy from a rotating black hole is the Blandford–Znajek mechanism \citep{1977MNRAS.179..433B}. Here, the magnetic field lines, anchored in the surrounding accretion disk, thread the event horizon of a black hole. The rotation of black hole 
%induces frame dragging within the ergosphere, twisting the magnetic field lines and 
generates an outward Poynting flux and this carries away the energy and angular momentum from the black hole, leading to an efficient transfer of spin energy into electromagnetic outflows. This can explain the power of the relativistic jets observed in AGN \citep{1984RvMP...56..255B} although the black hole spin 
measurements are challenging.
%, indirect estimates, theoretical models, and  observations of jet dynamics provide important constraints on the role of spin in AGN jet power.}
Nevertheless, the result obtained in this work again leaves an open question regarding the origin of the jet power. Probably, future studies of the source (particularly
at VHE energies) may provide better estimates for the hadron content in the jet. This will help to have a realistic estimate of the blazar jet
power which can be compared with the accretion power for more insights into the disk-jet connection in AGN.

%%%%%%%%%%%%%%%%%%%%%%%%%%%%%%%%%%%%%%%%%%%%%%%%%%%
%%%%%%%%%%%%%%%%%%%%%%%%%%%%%%%%%%%%%%%%%%%%%%%%%%%%%%%%
%%%%%%%%%%%%%%%%%%%%%%%%%%%%%%%%%%%%%%%%%%%%%%%%%%%
\begin{table*}
\centering
\renewcommand{\arraystretch}{1.6} 
    \resizebox{\linewidth}{!}{%
\begin{tabular}{lcccccccccc}
    \hline
    \hline
 %   \multicolumn{8}{c}{Broken Power-law}\\
    
    Period & p & q & $\gamma_{min}$ & $\gamma_{max}$ & $\gamma_b$  & $B$ & $\Gamma$  &$P_{jet,light}$& $P_{jet,heavy}$& ${\chi^2}/{dof}$
\\
    \hline

    State S1 & $2.16_{-0.11}^{+0.09}$  & $3.61_{-0.20}^{+0.21}$  & $4.0_{-0.56}^{+0.74}$ & $1.41\times 10^4$ & $544$  & $1.78_{-0.10}^{+0.10}$ & $32.94$   
    & $46.94$&$48.68$&$53.42/6$\\
    
     State S2  & $2.71_{*}^{+0.11}$  & $3.21_{-0.34}^{+0.36}$    & $4.0_{-0.53}^{+0.58}$ & $2.09\times 10^4$  & $237$  & $2.10_{-0.14}^{+0.14}$ & $15.02$ & $46.41$&$48.62$&$36.31/10$\\

   State S3 & $2.18_{-0.13}^{+0.10}$  & $3.35_{-0.21}^{+0.22}$    & $4.0_{-0.49}^{+0.57}$ & $1.73\times 10^4$ & $559$  & $1.88_{-0.16}^{+0.17}$& $28.92$ 
   &$46.80$&$48.59$&$46.48/11$\\
   
   \hline
   \hline
  \end{tabular}
}
 % \vspace{0.5cm}
\caption{Details of the ﬁt parameters obtained by ﬁtting the chosen states S1, S2 and S3 using one zone leptonic model. Column description; 1. Diﬀerent ﬂux states; 2, 3. Particle BPL spectral indices (p, q); 4, 5, 6. Minimum, maximum electron energy and break energy ($\gamma_{min}, \gamma_{max}, \gamma_{b} $); 7. Magnetic ﬁeld (B) in units of G; 8. Bulk Lorentz factor ($\Gamma$); 9, 10. Logarithmic light jet and heavy jet kinetic powers ($P_{jet,light}$,$ P_{jet,heavy}$) in units of $erg\,s^{-1}$; 11. ${\chi^2}/{dof}$ for a particular fit;. The viewing angle of the jet is 3 degree. The subscript and superscript values on parameter are lower and upper error values of model parameters, respectively, obtained through spectral fitting. Here, $*$ implies that the lower or upper error on the parameter is not constrained.}
    \label{tab:Brd_SED_parms} 
\end{table*}
%%%%%%%%%%%%%%%%%%%%%%%%%%%%%%%%%%%%%%%%%%%%%%%%%%%

%. %%SED figxxin%%%%%%%%onding %%%%%%%%%%%%%%%%%%%%%%%%%%%%%%%%%%%%%%%%%%%%%
%%%%%%%%%%%%%%%%%%%%%%%%%%%%%%%%%%%%%%

%%%%%%%%%%%%%%%%%%%%%%%%%%%%%%%%%%%%%%%%%%%%%%%%%%%%%%%%%%%%%%%%%%%%%%%%%%%%%%%%%%%%%%%%%%%%%%%%%%%%%%%
\section{SUMMARY}
\label{sec:sum}

The continuous observation by \emph{Fermi}-LAT, combined with simultaneous Swift-XRT/UVOT observations, provided a unique opportunity to study the broad-band temporal and spectral behavior of the FSRQ type blazar B2\,1348+30B. The 3\,day binned $\gamma$-ray lightcurve revealed one quiescent and two active states with simultaneous X-ray and optical/UV observations. A detailed spectral and temporal analysis was performed for these three selected states which are labelled as S1 (minor flare), 
S2 (quiescent), and S3 (major flare). The summary of our detailed analysis are:
%The continuous observation of the FSRQ B2\,1348+30B by \emph{Fermi}-LAT, coupled with simultaneous observations from Swift-XRT/UVOT, has first time provided an opportunity to study the  broad-band temporal and spectral behavior of the source. The 3-day bin $\gamma$-ray lightcurve of the source revealed  one quiescent and two active states having simultaneous observations in optical/UV bands, respectively(see Fig.\,\ref{fig:3_day_fermi}). We performed a detailed spectral and temporal analysis of the source during the three selected states S1, S2 and S3. The summery of the present work is as follows
\begin{itemize}
    \item  The Bayesian block analysis approach suggests the minimum variability timescale of the source can be considered  $\leq$ 3 days. From the light travel time arguments, the estimated emission region size is $R \leq 9\times10^{16}\,cm$ and the distance of the emission region from the central black hole is $\approx 3.7\times 10^{16}cm$, by assuming $\delta\approx 20$.
    \item The source shows significant variability in the optical and $\gamma$-ray bands, likely due to high-energy electrons, while the UV and X-ray bands exhibit moderate variability, attributed to lower-energy electrons. Broad-band spectral modeling supports this, with high-energy electrons contributing more to the variability. Interestingly, the UV band displays reduced variability, which may be due to a substantial non-variable contribution (e.g. thermal emission from the accretion disk).

 %    The source exhibited significant variability in optical and $\gamma$-ray bands, attributed to high-energy electrons, while X-ray and UV bands showed moderate variability, likely due to lower-energy electrons. Interestingly, the UV band exhibits lower variability, potentially due to a significant non-variable accretion disk contribution. This possibly suggests a combined influence of both the jet and the disk on the low-energy emissions. This result is consistent with the UVOT spectral analysis of the source.
    
    \item The $\rm TS_{curve}$ values obtained for the states S1, S2, and S3 indicate that the $\gamma$-ray spectra of the source are best fitted with a power-law model. Additionally, no photons with energies $\geq$ 20 GeV and significance $\geq3\sigma$  were detected from the source.
    
    \item The broken power-law fit for the X-ray spectra of the state S2 indicates this energy band falls in the transition regime between the synchrotron and 
	Compton spectral components. The high energy index of the broken power-law is very hard ($\sim$ 0.8) and can be attributed to the low-energy cutoff in the 
	underlying electron distribution.
	%providing constraints on the jet kinetic power.
	%, indicating that the spectrum lies between the synchrotron and Compton spectral components. This hard X-ray spectrum 
    
    %The slope of the X-ray continuum offers valuable insights into the acceleration mechanisms governing relativistic electrons. Spectral indices ($\Gamma$ < 1.5) from power-law fits across all three states suggest that the X-ray emission primarily originates from the Compton process rather than synchrotron radiation, as it aligns with the rising part of the Compton spectrum.
   % In particular, the hard X-ray spectrum observed in state S2 ($\Gamma \sim$ 0.8) corresponds to a particle index of 0.6 under the Thomson cross-section. This deviates from what is expected in standard Fermi shock acceleration, pointing to the potential existence of a $\gamma_{min}$ cutoff in the Compton component of the SED.
    
    \item The broad-band SED can be well reproduced under one-zone leptonic model involving synchrotron, SSC and EC/IR emission processes.
We found nearly a twofold increase in the bulk Lorentz factor during flaring states.
    
    \item The hard X-ray spectra of the state S2 led us to estimated the minimum energy of the emitting electron distribution as $\gamma_{min}=4$.
		Combining this with the best fit parameters obtained from the broadband SED modelling, we found the limits on the blazar jet power as
		$2.57\times 10^{46}$ erg/s (for light jet) and $4.17\times 10^{48}$ erg/s (for heavy jet). This power is significantly larger than the Eddington power estimated from the 
		blackhole mass and this suggests the blazar jet may extract power from the spin of the black hole.
    
\end{itemize}

%\newpage

\section*{Acknowledgements}

We thank the anonymous referee for constructive comments and suggestions. The authors thank the Department of Physics, University of Kashmir for providing the necessary facilities to carry out this research work. ZS is supported by the Department of Science and Technology (DST), Govt. of India, under the INSPIRE Faculty grant (DST/INSPIRE/04/2020/002319). SA also thanks IUCAA for providing the facilities. This research has made use of $\gamma$-ray data from \emph{Fermi} Science Support Center (FSSC). The work has also used the Swift Data from the High Energy Astrophysics Science Archive Research Center (HEASARC), at NASA’s Goddard Space Flight Center.

%%%%%%%%%%%%%%%%%%%%%%%%%%%%%%%%%%%%%%%%%%%%%%%%%%
\section*{Data Availability}

The data and the model used in this article will be shared on
reasonable request to the corresponding author, Sajad Ahanger (email: sajadphysics21@gmail.com)

%%%%%%%%%%%%%%%%%%%% REFERENCES %%%%%%%%%%%%%%%%%%

% The best way to enter references is to use BibTeX:

\bibliographystyle{harv}

% Loading bibliography database
\bibliography{ref}

% Biography
\bio{}
% Here goes the biography details.
\endbio

\bio{}
% Here goes the biography details.
\endbio

\end{document}